\documentclass[useAMS,referee]{mn2e}
\usepackage{graphicx}
\usepackage{amsmath}

\title[SWIFT J0513.4--6547]{ \emph{RXTE} \& \emph{Swift} Observations of SWIFT J0513.4--6547}
\author[\c{S}. \c{S}ahiner, M. M. Serim, A. Baykal, and S. \c{C}. \.{I}nam ]
{\c{S}. \c{S}ahiner$^{1}$\thanks{E-mail: seyda@astroa.physics.metu.edu.tr (\c{S}\c{S}); muhammed@astroa.physics.metu.edu.tr (MMS);
 inam@baskent.edu.tr (S\c{C}\.{I}); altan@astroa.physics.metu.edu.tr (AB)}, M. M. Serim$^{1}$\footnotemark[1], 
 A. Baykal$^{1}$\footnotemark[1] and S. \c{C}. \.{I}nam$^{2}$\footnotemark[1] \\
$^{1}$Physics Department, Middle East Technical University, 06531 Ankara, Turkey\\
$^{2}$Department of Electrical and Electronics Engineering, Ba\c{s}kent University, 06810 Ankara, Turkey}

\begin{document}

\date{Received 2015}

\pagerange{\pageref{firstpage}--\pageref{lastpage}} \pubyear{2015}

\maketitle

\label{firstpage}

\begin{abstract}

We present timing and spectral analysis of  \emph{Swift}$-$XRT and \emph{RXTE}$-$PCA observations of the transient Be/X-ray pulsar SWIFT J0513.4--6547 during its outburst in 2009 and its rebrightening in 2014. From 2009 observations, short term spin-up rate of the source after the peak of the outburst is found to have about half of the value measured at the peak of the outburst by Coe et al. When the source is quiescent between 2009 and 2014, average spin-down rate of the source is measured to be $\sim 1.52 \times 10^{-12}$ Hz s$^{-1}$ indicating a surface dipole magnetic field of $\sim 1.5 \times 10^{13}$ Gauss assuming a propeller state. From 2014 observations, short term spin-down rate of the source is measured to be about two orders smaller than this long-term spin-down rate. The orbit of the source is found to be circular which is atypical for transient Be/X-ray binary systems.  Hardness ratios of the source correlate with the X-ray luminosity up to $8.4\times 10^{36}$ erg s$^{-1}$ in 3-10 keV band, whereas for higher luminosities hardness ratios remain constant.  Pulsed fractions are found to be correlated with the source flux. Overall \emph{Swift}$-$XRT and \emph{RXTE}$-$PCA energy spectrum of the source fit equally well to a model consisting of blackbody and power law, and a model consisting of a power law with high energy cut-off. From the pulse phase resolved spectra and pulse phase resolved hardness ratios obtained using \emph{RXTE}$-$PCA, it is shown that spectrum is softer for the phases between the two peaks of the pulse.   

\end{abstract}


\section{Introduction}

On 2009 April 2 (MJD 54923) the \emph{Swift}$-$BAT (Burst Alert Telescope) detected a new transient X-ray source 
SWIFT J0513.4--6547 in the Large Magellanic Cloud (LMC) (Krimm et al. 2009). An archival analysis of BAT data brought out 
that the source had been active since 2009 March 4 (MJD 54894) and had reached a peak brightness of $\sim40$ mCrab on 
2009 March 6. Monitoring observations with \emph{RXTE} (\emph{Rossi X-ray Timing Explorer}) tracked the decline of the 
outburst until 2009 May 13 (MJD 54964). The detection of strong pulsations with a period around 27.28 s during the outburst 
and the association of the X-ray transient with the optical source 2MASS 05132826--6547187, have led SWIFT J0513.4--6547 to be identified as a 
High Mass X-ray Binary system (HMXB) (Krimm et al. 2009, Krimm et al. 2013). 

A recent study (Coe, Udalski \& Finger 2013, Coe et al. 2015) on 4.5 years of OGLE IV (Optical Gravitational Lensing 
Experiment Phase IV) light curve revealed regular modulation of the optical magnitude. Approximately 0.05 magnitude flickering 
in every 27.4 days confirms the binary nature of SWIFT J0513.4--6547, assuming that the optical modulation represents 
the binary period of the system. Coe et al. (2015) refined the frequency measurements of Finger \& Beklen (2009) from 
\emph{Fermi}/GBM (Gamma Ray Burst Monitor) detections and derived the orbital parameters of the system. They provided an upper 
limit of 0.17 to the eccentricity using a model with a fixed orbital period of 27.405 d. 

Coe et al. (2015) also studied the optical spectrum and identified the companion as a B1Ve star. The position of 
SWIFT J0513.4--6547 on the Corbet diagram (pulse period vs. orbital period diagram, Corbet 1984) affirms the Be/X-ray binary 
nature. Generally X-ray outbursts of Be/X-ray binary systems occur in two different forms, classified as Type I or Type II 
(Stella, White \& Rosner 1986). Type I outbursts ($L_{x} \sim 10^{36-37}$ erg s$^{-1}$) are correlated with the binary orbit 
so that; they are recurrently observed in certain orbital phases due to enhanced mass transfer. In contrast, Type II 
outbursts ($L_{x} \geq 10^{37}$ erg s$^{-1}$) are less frequent, irregular but giant eruptions that decay along several 
orbital periods. 2009 outburst of SWIFT J0513.4-6547 can be identified as Type II since its peak luminosity is derived as 
$1.3 \times 10^{38}$ erg s$^{-1}$ (Coe et al. 2015) and the outburst lasted for about two orbital cycles. 

Recently, the source underwent another bursting activity in 2014. During the \emph{XMM-Newton} 
(\emph{X-ray Multi-Mirror Mission}) detection on 2014 August 25 (MJD 56894), the luminosity was at the order of 
$10^{36}$ erg s$^{-1}$ (Sturm et al. 2014). Seven follow-up \emph{Swift} observations until the end of the year show 
that the luminosity peaked twice at the same level, September 19 (MJD 56919) and October 16 (MJD 56946). These three 
outbursts on August 25, September 19 and October 16 can be classified as Type I since their occurrence times are just about the expected optical maxima. 

In this paper, we study the spectral and timing properties of SWIFT J0513.4--6547 through its observations with \emph{RXTE} 
and \emph{Swift} during the 2009 outburst. In addition, we analyse the \emph{Swift} data of 2014 outburst. We 
introduce the observations in Section 2. We present our timing and spectral results in Sections 3 and 4. In Section 5, we 
discuss our results and conclude.

\section{Observations}

\subsection{\emph{RXTE}}

\begin{table}
  \caption{Log of \emph{RXTE}$-$PCA observations of SWIFT J0513.4--6547.}
  \label{rxteobs}
  \center{\begin{tabular}{cccc}
  \hline	
\emph{RXTE} 	& Time 	& Number of 	& Exposure \\
proposal ID 	& (MJD) & observations 	& (ks) \\	
 \hline
 94065	& 54935 - 54947 & 10 & 14.24 \\
 94421	& 54948 - 54964 & 8  & 9.73 \\
\hline
\end{tabular}} \\
\end{table} 

18 target of opportunity (TOO) observations (see Table \ref{rxteobs}) with \emph{RXTE}$-$PCA (Proportional Counter Array; 
Jahoda et al. 1996) were carried out between 2009 April 14 (MJD 54935) and 2009 May 13 (MJD 54964). These observations 
witness the decline of the 2009 outburst of SWIFT J0513.4--6547. The data are subjected to standard analysis using the tools 
of \verb"HEASOFT V.6.15.1". Time filtering is applied with the criteria on elevation angle to be greater than 10$\degr$, 
offset from the source to be less than 0$\degr$.02 and electron contamination of PCU2 to be less than 0.1. We also exclude 
data within 30 minutes of the peak of South Atlantic Anomaly (SAA) in order to increase signal-to-noise ratio (SNR). The 
resulting net exposure after the filtering is 24 ks. 

GoodXenon events are used to generate 0.1 s binned 3--10 keV light curves for the timing analysis, whereas Standard2f mode 
data are considered for broad spectral analysis. Furthermore, using \verb"FASEBIN" tool, we construct pulse phase resolved 
spectra from GoodXenon events. Energy-resolved pulse profiles are produced from phase spectra by obtaining the count rates 
per phase bin with the tool \verb"FBSSUM". We select only data from PCU2 during the extraction of spectral files, since it 
is the only PCU that is active during all of the observations. Moreover, only photons detected on the top anode layers are 
selected to increase SNR. Background subtraction is applied for spectra and light curves by using EPOCH 5C background model 
supplied for faint sources (i.e. $\textless$ 40 cts/s/PCU). 

\subsection{\emph{SWIFT}}

\begin{table}
  \caption{Log of \emph{Swift}$-$XRT observations of SWIFT J0513.4--6547. The observations given above the horizontal line are
  from 2009, whereas 2014 observations are below the line.}
  \label{swiftobs}
  \center{\begin{tabular}{cccc}
  \hline	
 \emph{Swift}	& Time 	& Off-axis 	& Exposure \\
 obsID 		& (MJD) & ($\prime$) 	& (ks) \\	
 \hline
 00031393001  & 54932 & 0.45 & 4.63 \\
 00031393002  & 54937 & 0.35 & 6.34 \\
 00031393003  & 54940 & 1.01 & 1.69 \\
 \hline
 00031393004  & 56916 & 0.10 & 0.93 \\
 00031393005  & 56919 & 2.12 & 1.41 \\
 00031393006  & 56925 & 0.39 & 0.97 \\
 00031393007  & 56939 & 2.48 & 1.00 \\
 00031393008  & 56946 & 1.53 & 1.24 \\
 00033532001  & 56985 & 2.39 & 1.31 \\
 00033532002  & 56986 & 1.21 & 1.05 \\
 \hline
\end{tabular}} \\
%
\end{table} 

After the \emph{Swift}$-$BAT discovery of SWIFT J0513.4--6547, 3 follow-up \emph{Swift}$-$XRT (X-ray Telescope; 
Burrows et al. 2005) observations were performed on 2009 April. Moreover; when the source re-brightened on 2014 August, 
7 pointing observations were completed until the end of 2014. A log of \emph{Swift}$-$XRT observations is given in 
Table \ref{swiftobs}. The total exposure of 2009 and 2014 observations are 12.67 ks and 7.92 ks, respectively. 
The observations of SWIFT J0513.4--6547 are in photon counting (PC) mode which has a time resolution of $\sim$2.5 s. 
XRT event files are filtered with the default screening criteria of \verb"XRTPIPELINE V.0.13.0" script of 
\verb"XRTDAS V.2.9.2" package. Standard grade filtering (0-12) is applied to the data. For each observation, vignetting 
corrected exposure maps are generated during the pipeline processing. 

Throughout the 2009 observations of SWIFT J0513.4--6547 count rates were just around the pile-up limit 
($\sim$ 0.5 cts s$^{-1}$), therefore the events are corrected for pile-up. For these observations an annular source region 
is selected in order to exclude an inner circle of radius $\sim$7 arcsec, while the outer radius of the annulus is 
$\sim$70 arcsec. In contrast, the count rates of 2014 observations were low ($\sim$ 0.1 cts s$^{-1}$), consequently a 
circular source extraction region with $\sim$35 arcsec radius is chosen. The same circular source-free region with 
$\sim$140 arcsec radius is used for background extraction for both 2009 and 2014 observations. Light curves and spectra are 
extracted by filtering selected regions using \verb"XSELECT V.2.4C".

For the overall spectral analysis spectral files are rebinned to have at least 20 counts per energy bin, which is required 
for $\chi^2$ fitting. However, spectra of individual observations are rebinned to have minimum of 1 count per bin and Cash 
statistic (Cash 1979) is used as it is advised for low count spectra. We use the latest response matrix file (version v014) and 
create individual ancillary response files using the tool \verb"XRTMKARF V.0.6.0" with the exposure maps produced during 
the pipeline processing. Spectral analysis is performed using \verb"XSPEC V.12.8.1g".  

\section{Timing Analysis}

\subsection{Pulse Arrival Time and Pulse Frequency Measurements}

For the \emph{RXTE}$-$PCA and \emph{Swift}$-$XRT observations in 2009 (see Tables \ref{rxteobs} and \ref{swiftobs}),  we construct background subtracted lightcurves. The resulting \emph{RXTE} and \emph{Swift} lightcurves are 0.1s and 2.51s binned respectively. 
We correct the lightcurves with respect to the Solar System barycenter. The lightcurves are also corrected for the orbital binary motion using the orbital parameters given by Coe et al. (2015).   
 
The nominal pulse period is estimated, by folding \emph{RXTE}$-$PCA
barycentric lightcurve on statistically independent trial periods (Leahy et al. 1983). We search for periodicity using $\chi^{2}$ test and use the template pulse profile giving the maximum $\chi^{2}$. Then we fold the \emph{RXTE}$-$PCA and \emph{Swift}$-$XRT lightcurves with the best period obtained from  \emph{RXTE}$-$PCA lightcurve. Pulse profiles are obtained with 10 and 20 phase bins from \emph{Swift}$-$XRT 
and \emph{RXTE}$-$PCA, respectively. Pulse profiles are represented by their Fourier harmonics 
(Deeter \& Boynton 1985). Then, we obtain the pulse arrival times for each observation by cross correlating the pulses with template pulse profile.

\begin{figure}
  \center{\includegraphics[width=9.5cm,angle=0]{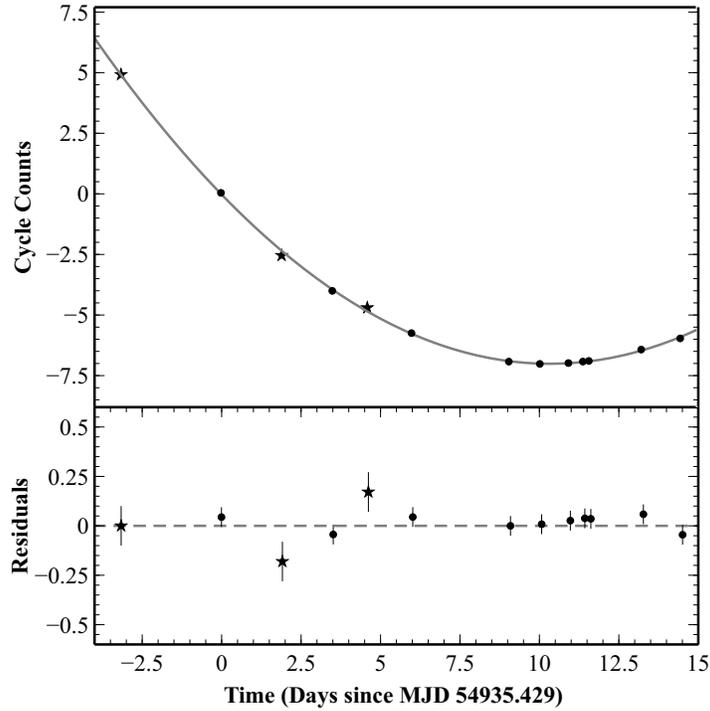}} 
  \caption{{\bf{(top)}} Arrival times obtained from \emph{RXTE}$-$PCA and \emph{Swift}$-$XRT observations in 2009 after binary orbital correction. Data points indicated by asterisks correspond to \emph{Swift}$-$XRT measurements. Quadratic fit is shown as the grey line. {\bf{(bottom)}} Residuals after the quadratic trend is removed.}
  \label{arrival}
\end{figure}

The phase connected pulse arrival times
 after the binary orbital correction are shown in the top panel of Fig. \ref{arrival}, whereas the residuals after the removal of the quadratic trend are shown in the bottom panel of Fig. \ref{arrival}. From this quadratic trend, spin-up rate is found to be $1.74(6) \times 10^{-11} $Hz s$^{-1}$. Reduced chi square of this fit is around 1.07. This suggest that the orbit of the system is circular which was also suggested by Coe et al. (2015). Coe et al. (2015) also found an upper limit of 0.17 to the eccentricity within $2 \sigma$ confidence level. In order to find a further constraint on eccentricity,
an elliptical orbital model is fitted to the residuals as suggested by Deeter et al. (1981).
This fit gives an upper limit to the eccentricity as 0.028 in $2 \sigma$ level.
It should also be noted that the time coverage of pulse arrival times is only 63 percent of orbital period. Full orbital coverage of arrival times is needed to obtain a better constraint on the eccentricity.

\begin{figure}
  \center{\includegraphics[width=9.5cm,angle=0]{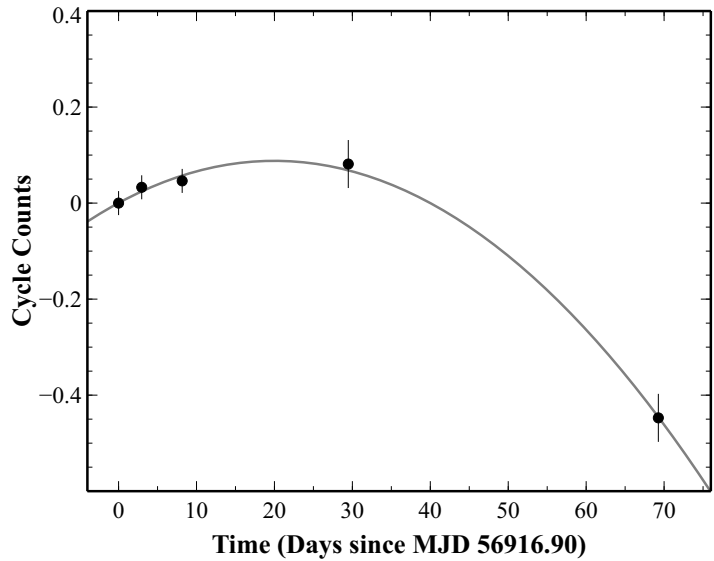}} 
  \caption{Arrival times obtained from  \emph{Swift}$-$XRT observations in 2014 after binary orbital correction. Quadratic fit is shown as the grey line.}
  \label{arrival2014}
\end{figure}

SWIFT J0513.4-6547 was also observed seven times in 2014 within a time span of $\sim 70$ days (see Table \ref{swiftobs}). As for the 2009 observations, we construct background subtracted 2.51s binned lightcurves from 2014 observations. 
We construct pulse phase profiles for five of these observations and connect them in phase.

 Pulse arrival times are presented in Fig. \ref{arrival2014}. Among 2014 observations listed in Table \ref{swiftobs}, no significant pulse detection is obtained from the observation at MJD 56939, and a single pulse arrival time is measured from the observations at MJD 56985 and MJD 56986. The pulse frequency measurement of the timing solution is consistent with the value reported by Sturm et. al. (2014) within one sigma level. Furthermore a spin down rate of $(5.92 \pm 1.9) \times 10^{-14}$  Hz s$^{-1}$ is found during this 70 days time span.
Timing solutions of 2009 and 2014 are given in Table \ref{timing}. 

\begin{table}
  \caption{Timing solutions of 2009 and 2014 observations.}
  \label{timing}
  \center{\begin{tabular}{ccc}
  \hline	 \hline
 {\bf{Parameter}}	& {\bf{2009 Solution}} 	& {\bf{2014 Solution}} \\
 \hline
Epoch (MJD) &  54935.42866 & 56916.89857 \\
$\nu$ ($10^{-2}$ Hz) & 3.67020(7) & 3.64432(5) \\
$\dot{\nu}$ (Hz s$^{-1})$ & $1.74(6) \times 10^{-11}$ & $-(5.92\pm 1.9) \times 10^{-14}$ \\
 \hline \hline
\end{tabular}} \\
\end{table}
 
\subsection{Energy-Resolved Pulse Profile}
 
\begin{figure*}
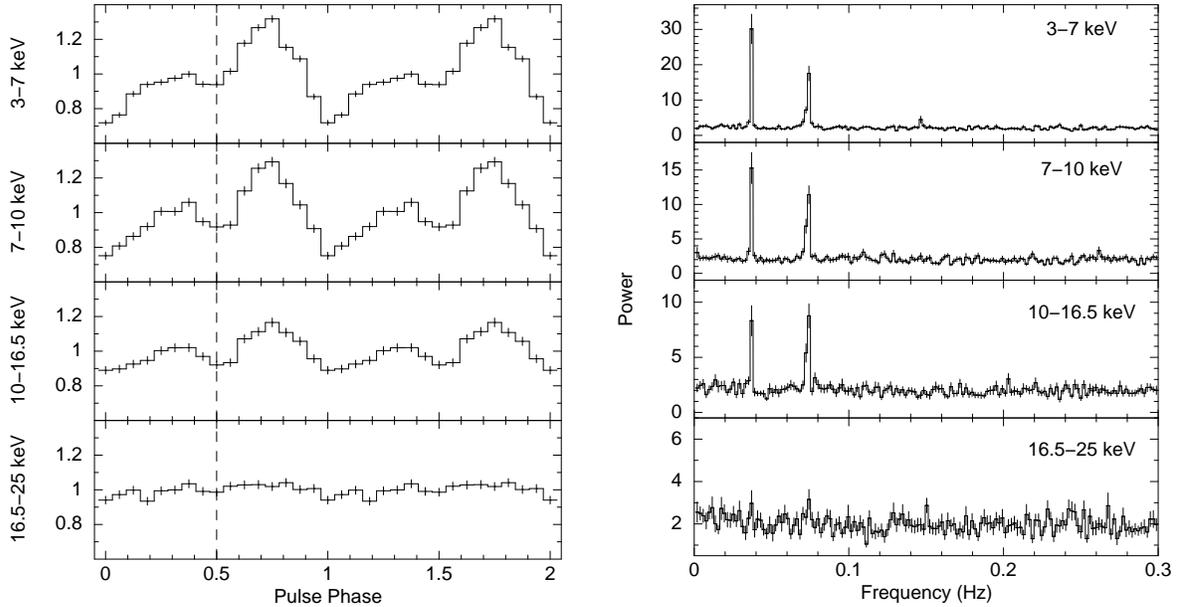

  \center{\includegraphics[width=8cm, angle=270]{all_norm2.eps}\includegraphics[width=8cm, angle=270]{pow_enres.eps}} 
  \caption{Energy resolved pulse profiles of \emph{RXTE}$-$PCA observations are given on the left. Y-axes are normalised 
  counts from the energy bands given on y-axes labels. The dashed line represents the local minima in between the double 
  peaks of the profiles. Energy resolved power spectra of the corresponding energy bands are given on the right.}
  \label{enres}
\end{figure*} 

To see if there are 
any energy dependent changes in the profile, we construct 3--7 keV, 7--10 keV, 10--16.5 keV  and 16.5--25 keV pulse profiles.  The energy-resolved profiles given in the left panels of Fig. \ref{enres} are profiles averaged for \emph{RXTE}$-$PCA data between 
MJD 54935--54950. These energy bands are selected according to their average count rates. The source is the strongest in 3--7 keV band with an average count rate of 3.2 cts s$^{-1}$. Other bands have approximately the same count rates as 1.7 cts s$^{-1}$ on average.  

Energy-resolved power spectra of the same energy bands are also constructed for comparison (see the right panels of 
Fig. \ref{enres}). The second harmonic (0.0734 Hz) of the pulse frequency, that indicates the double-peaked structure, is 
evident on power spectra. In 3--7 keV and 7--10 keV bands, pulse profiles and power spectra are similar. In 10--16.5 keV band, 
along with a significant drop in the pulsed fraction; the double-peaked structure strengthens. The count rate of the phase between the primary and the secondary peak (indicated with a dashed line in Fig. \ref{enres}) reaches the same level of the count rate of pulse minimum phase, which is the 0.0 phase.  Consequently, 
the powers of the pulse and its second harmonic are balanced on the 10--16.5 keV power spectrum. The pulse diminishes in the 
16.5--25 keV band, although the average count rate is similar to former bands. The powers of pulse and its second harmonic are at the level of noise on the 16.5-25 keV power spectrum.

\subsection{Hardness Ratio}

\begin{figure}
  \center{\includegraphics[width=10cm]{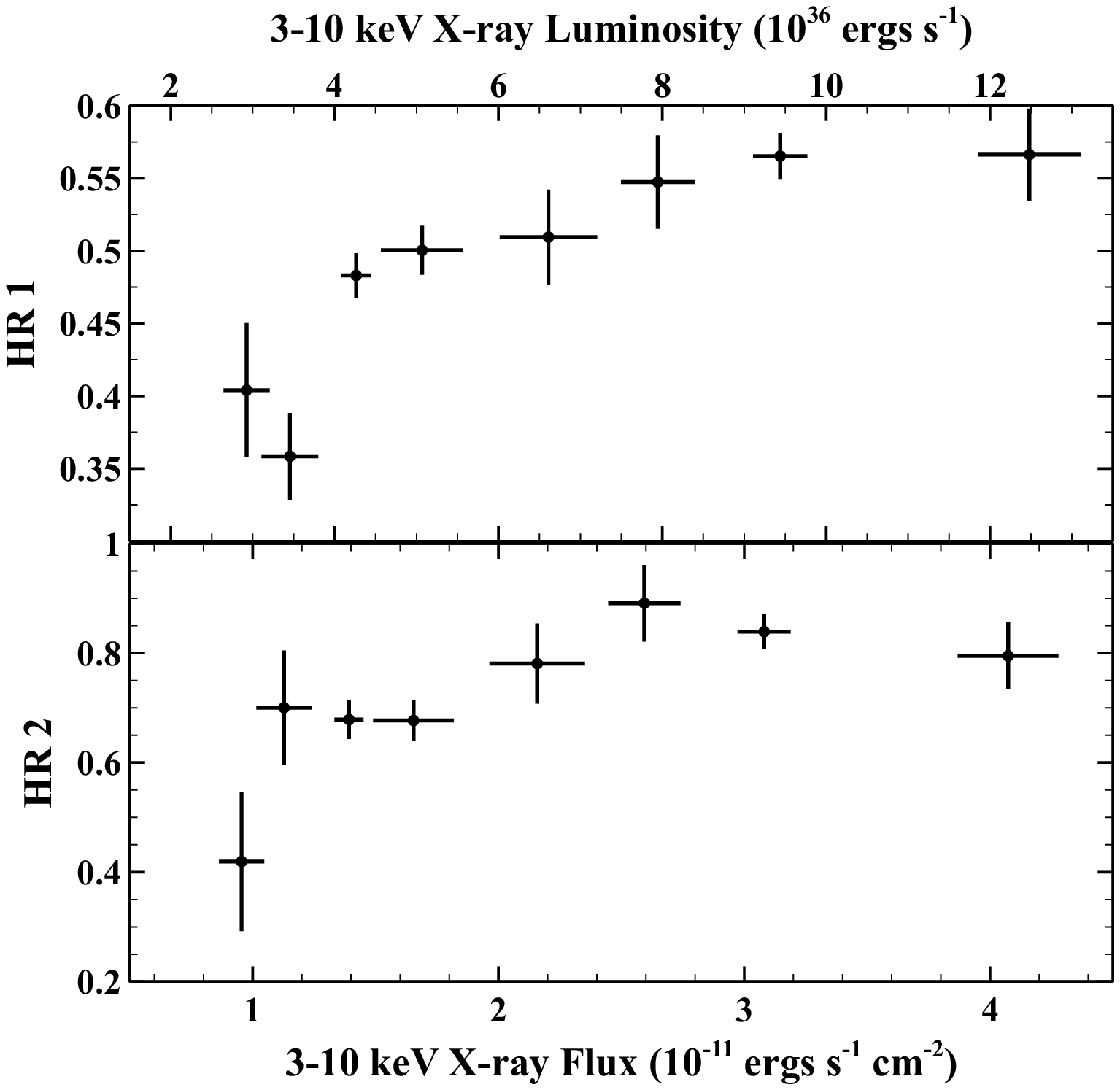}} 
  \caption{X-ray flux and X-ray luminosity dependences of the hardness ratios. HR1 (7--10 keV / 3--7 keV)and HR2 (10--16.5 keV/7--10 keV) ratios 
  are plotted over daily averaged 3--10 keV X-ray flux and X-ray luminosity values.}
  \label{hard}
\end{figure}

\begin{figure}
  \center{\includegraphics[width=8.0cm,angle=270]{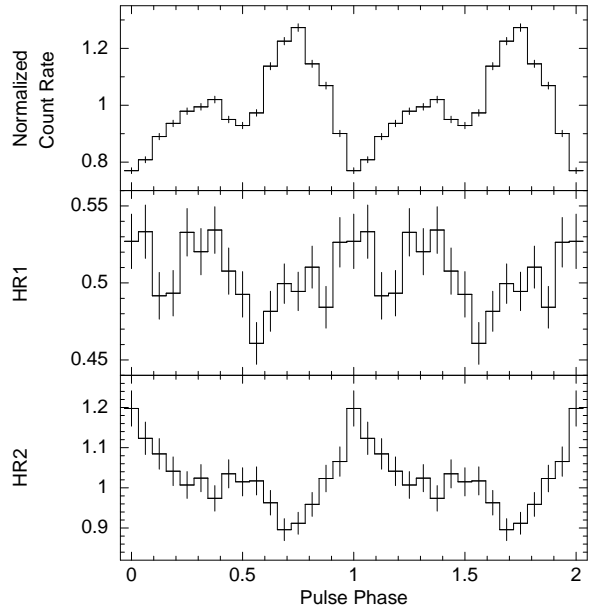}} 
  \caption{Pulse phase resolved hardness ratios. The 3--16.5 keV pulse profile is plotted on the top panel. 
  7--10 keV/3--7 keV and 10--16.5 keV/7--10 keV a count rate ratios are denoted as HR1, and HR2  on y-axis labels respectively.}
  \label{hrp}
\end{figure}

We also examine the lightcurves corresponding to specific energy bands and pulse profiles of SWIFT J0513.4--6547 in the context of hardness 
ratio. Daily averaged X-ray flux and X-ray luminosity dependences of HR1 (7--10 keV/3--7 keV) and HR2 (10--16.5 keV/7--10 keV) are plotted in 
Fig. \ref{hard}. 

Hardness ratios in pulse phase are plotted in Fig. \ref{hrp}. The top panel is the average pulse profile in the 3--16.5 keV 
energy range, which is the combination of pulsed bands in Fig. \ref{enres}. Hardness ratios HR1 and HR2 are calculated 
as un-normalized count rate ratios of consecutive bands in Fig. \ref{enres}. The HR1 reaches its minimum between the double 
peaks of the pulse. The secondary peak has the hardest ratio, whereas the primary peak has an average HR1 value. For HR2 the 
hardest phase is the 0.0 phase. A gradual softening is observed until the primary peak, where HR2 starts to increase again.

\subsection{Pulsed Fraction} 

\begin{figure}
  \center{\includegraphics[width=4.6cm,angle=270]{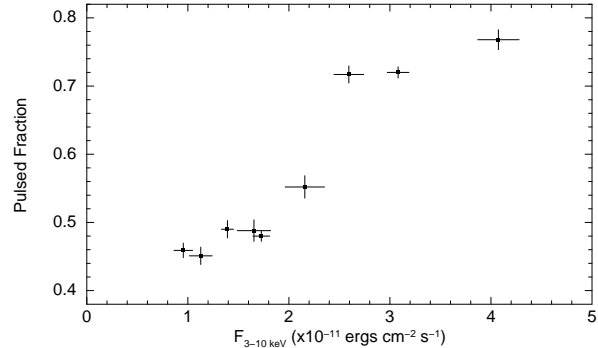}} 
  \caption{Pulsed fraction as a function of unabsorbed flux. Pulsed fractions are calculated from the count rates of
  phase-binned spectra, in the energy range 3--16.5 keV.}
  \label{pf}
\end{figure}

3--16.5 keV pulse profiles of individual \emph{RXTE}$-$PCA observations are constructed from the count rates of phase-binned 
spectra using the tool \verb"FBSSUM". Then, the pulsed fractions are calculated with the standard definition 
$\frac{p_{max}-p_{min}}{p_{max}+p_{min}}$; where $p_{min}$ and $p_{max}$ are minimum and maximum counts of the pulse. 

In Fig. \ref{pf}, measured pulsed fractions are plotted as a function of  3--10 keV unabsorbed flux values obtained from the spectral fits. A positive correlation between X-ray flux and pulsed fraction is evident.

\section{Spectral Analysis}

\subsection{Overall Spectrum during 2009}

\begin{table}
  \caption{Best fit spectral parameters for \emph{Swift} 2009, \emph{Swift} 2014 and for the simultaneous fitting of 
  \emph{Swift} and \emph{RXTE} data from 2009 (see Fig. \ref{simspe}). All uncertainties are calculated at the 
  90 per cent confidence level. The acronyms used for the different model components are: $\it{ph}$ for \texttt{PHABS}, 
  $\it{po}$ for \texttt{POWERLAW}, $\it{bb}$ for \texttt{BBODYRAD} and $\it{hi}$ for \texttt{HIGHECUT}.}
  \label{spepar}
  \center{\renewcommand{\arraystretch}{1.5}\begin{tabular}{lccc}
  \hline \hline 
 \emph{Swift} \& \emph{RXTE} 2009 & \it{ph*po} &  \it{ph*(bb+po)} & \it{ph*po*hi} \\
 \hline 
 $n_{\mathrm{H}}$ $[10^{21}\,$cm$^{-2}]$	& $2.58^{+0.33}_{-0.30}$ & $1.66^{+0.29}_{-0.26}$ & $1.20^{+0.24}_{-0.22}$ \\
 $\Gamma$ 			& $1.32\pm0.04$ 	 & $1.32\pm0.06$ 	  & $0.93\pm0.04$ \\	
 $kT_{\mathrm{BB}}$ [keV]		& -- 			 & $1.89\pm0.11$ 	  & -- \\
 $R\,^{(a)}$ [km]		& --			 & $1.73\pm0.17$ 	  & -- \\
 $E_{\mathrm{cut}}$ [keV]		& --			 & --			  & $5.04^{+0.77}_{-0.67}$ \\
 $E_{\mathrm{fold}}$ [keV]		& --			 & --			  & $11.59^{+1.67}_{-1.56}$ \\
 Flux$\,^{(b)}$ 		& $3.95\pm0.15$ 	 & $4.06\pm0.16$ 	  & $3.93\pm0.14$ \\
 \hline
 $\chi^{2}_{\mathrm{red}}\,$/$\,$dof	& 1.53/186		 &  0.96/184		  & 0.96/184 \\
 \hline\hline
  \end{tabular}
  \begin{tabular}{lcccc}
  & & & & \\
 & $n_H$ & $\Gamma$ & Flux$\,^{(b)}$ & $\chi^{2}_{Red}\,$/$\,$dof \\
\hline
 \emph{Swift} 2009 & $1.35^{+0.31}_{-0.29}$ & $0.99\pm0.07$ & $4.2\pm0.16$ & 0.98/155  \\
 \emph{Swift} 2014 & $1.32^{+0.70}_{-0.58}$ & $1.13^{+0.17}_{-0.16}$  &  $0.8\pm0.07$ & 0.98/32  \\
 \hline \hline
\end{tabular}} \\
\begin{flushleft}
$\bf{Notes.}$ $^{(a)}$ Radius of the blackbody emitting region calculated from the normalization parameter of 
\verb"BBODYRAD" model, assuming a 50.6 kpc distance to the LMC. $^{(b)}$ Unabsorbed flux values calculated using 
\verb"CFLUX" model for the energy range 0.3--10 keV, in units of $10^{-11}\,$erg$\,$cm$^{-2}\,$s$^{-1}$. 
\end{flushleft}
\end{table} 

The 0.3--10 keV \emph{Swift}$-$XRT spectrum of SWIFT J0513.4--6547 was previously described by a simple absorbed power law 
model using the 2009 observations of the source (Krimm et al. 2009, Coe et al. 2015). Our reanalysis of the \emph{Swift} 
2009 spectrum alone gives consistent results, however a trial for a simultaneous fit of \emph{Swift}$-$XRT and \emph{RXTE}$-$PCA 
data from 2009 showed that this basic model is not enough to fully describe the spectrum of SWIFT J0513.4--6547 
($\chi^{2}_{\mathrm{red}}\,$=1.53, see Table \ref{spepar} for best fit parameters). Energy ranges selected during the simultaneous 
fit of 12.7 ks XRT and 17.9 ks PCA spectra are 0.3--9.5 keV and 3--16.5 keV respectively. We exclude the energies above 
16.5 keV, due to the low statistical significance of the PCA spectral bins. A constant factor is involved in the model, in 
order to mimic the normalization uncertainty between the two instruments. 

\begin{figure}
  \center{\includegraphics[width=5.9cm, angle=270]{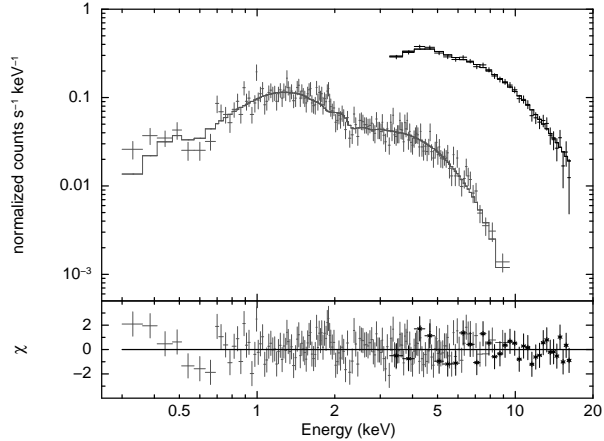}} 
  \caption{Simultaneous spectral fitting of \emph{Swift}$-$XRT (0.3--9.5 keV; grey) and \emph{RXTE}$-$PCA (3--16.5 keV; black) 
  data from 2009. The data and its best fit with \texttt{PHABS*POW*HIGHECUT} ($\chi^{2}_{Red}=0.98\,$; solid line) are 
  shown in upper panel, the residuals are given in the lower panel. See Table \ref{spepar} for the spectral parameters.}
  \label{simspe}
\end{figure}

We make an effort to improve the simultaneous spectral fit by including extra model components. We achieve similar 
improvements by adding a blackbody component (\verb"BBODYRAD") or a high energy cut-off (\verb"HIGHECUT"). The best fit 
plot is given in Fig. \ref{simspe} and the resulting best fit parameters are listed in Table \ref{spepar}. Both models result in an acceptable $\chi^{2}_{\mathrm{red}}$ (=0.96).

\subsection{Individual Spectra of 2009 observations}

\begin{figure}
  \center{\includegraphics[height=8.4cm, angle=270]{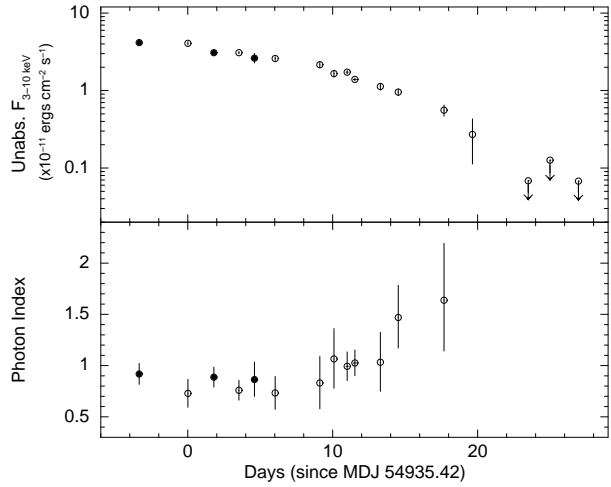}} 
  \caption{3--10 keV unabsorbed flux (upper panel) and photon index (lower panel) measurements for the individual 
  \emph{Swift}$-$XRT (filled circles) and \emph{RXTE}$-$PCA (empty circles) spectra from 2009. Error bars indicate 
  uncertainties at 90 per cent confidence level.}
  \label{spe2009}
\end{figure}

During 2009, a total of 21 individual observations (18 \emph{RXTE} and 3 \emph{Swift}) of SWIFT J0513.4--6547 cover a time 
span of about $\sim$30 days. We investigate the change of spectral parameters throughout this time by fitting the best fit 
model (\texttt{PHABS*POW*HIGHECUT}) to the individual spectra of each observation. The spectra of \emph{RXTE} observations 
which are on the same days are summed for a daily spectrum. SNR of the individual spectra are low due to both low exposure 
times ($\sim$ 1-2 ks) and low count rates, therefore we fix the cut-off model parameters (i.e. $E_{\mathrm{cut}}$ and $E_{\mathrm{fold}}$) 
to their best fit values given in Table \ref{spepar}. Moreover, the hydrogen column density ($n_H$) is preset to its best 
fit value, since we examine that the low $n_H$ along the line of sight does not show any significant change. Therefore we 
only evaluate the variability of photon index and unabsorbed flux. 3-10 keV fluxes are measured using the \verb"CFLUX" 
model in \verb"XSPEC". 

Daily spectral results for 3 XRT and 14 PCA spectra are plotted over observation mid-time in Fig. \ref{spe2009}. The upper 
panel shows the decline of the 2009 outburst of SWIFT J0513.4--6547. For observations after MJD 54955, we are only able to 
measure an upper limit to the flux. The lower panel of Fig. \ref{spe2009} demonstrates the increase in photon index by 
decreasing flux. The photon indexes measured from \emph{Swift} (filled circles) imply a slightly softer spectra when compared 
to values obtained from \emph{RXTE} (empty circles), however this might be an instrumental effect due to 
the different energy bandpass of detectors. In Fig. \ref{spe2009}, it is observed that as the flux decreases photon index increases which indicates that 
softer photons emerge from the source as the X-ray flux reduces.

\subsection{Re-brightening in 2014}

\begin{figure}
  \center{\includegraphics[width=3.5cm, angle=270]{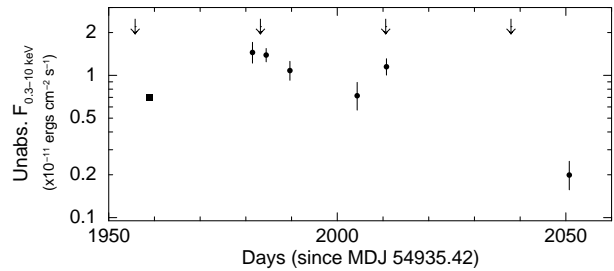}} 
  \caption{Unabsorbed flux (0.3--10 keV) values obtained from the individual spectral fits to the \emph{Swift} 2014 
  observations. The first data point at day $\sim1960$ is the \emph{XMM} flux measurement of Sturm et al. (2014). 
  The arrows above show the times of optical maximum according to the ephemeris given in Coe et al. (2015).
  Error bars indicate uncertainties at 90 per cent confidence level.}
  \label{2014flux}
\end{figure}

The 2014 brightening of SWIFT J0513.4--6547 is sparsely monitored by seven pointing \emph{Swift} observations (each $\sim$1 ks) 
between MJD 56916 and 56986.  The overall spectrum produced from 2014 observations provide similar results with \emph{Swift} 
2009 spectrum when described by a simple absorbed power law model (see Table \ref{spepar}). The absorption along the line of 
sight is estimated just the same with 2009 whereas spectral index appear to be a little higher, with a higher uncertainty 
due to the statistical quality of the low count spectrum. Nevertheless, the higher photon index in 2014 may also arise from 
the fact that the average X-ray flux of the source is about 20 per cent of the average 2009 flux, since we observe a flux dependence 
of the photon index for SWIFT J0513.4--6547 (see Fig. \ref{spe2009}). 
0.3-10 keV unabsorbed flux measurements of individual \emph{Swift} observations in 2014 are given in Fig. \ref{2014flux}. 
In this plot, the \emph{XMM-Newton} measurement on MJD 56894 is also shown (Sturm et al. 2014). Sturm et al. (2014) 
suggested that the 2014 brightening of SWIFT J0513.4--6547 might be explained by recurrent Type I outbursts, since the 
detections with \emph{XMM-Newton} and \emph{Swift} are close to the time of expected optical maxima. Therefore, we 
calculate consequent optical maxima using the ephemeris given in Coe et al. (2015) and indicate them with arrows on 
Fig. \ref{2014flux}. It is hard to distinguish whether the brightening is due to a single Type II outburst or a series of 
Type I outbursts, with the very few data points. One more supporting evidence for the Type I scenario is that, we measure 
another flux maximum corresponding exactly to the next optical maximum on MJD 56946. Furthermore, using the LMC distance of 
50.6 kpc (Bonanos et al. 2011), the peak flux on 2014 correspond to a luminosity of $4.3 \times 10^{36}$ erg s$^{-1}$ which is about two orders of 
magnitude lower than the maximum luminosity on 2009. The order of the X-ray luminosity also supports the Type I scenario for 
the 2014 brightening of SWIFT J0513.4--6547. 

\subsection{Pulse Phase Resolved Spectra}
 \begin{figure}
   \center{\includegraphics[width=9.65cm, angle=270]{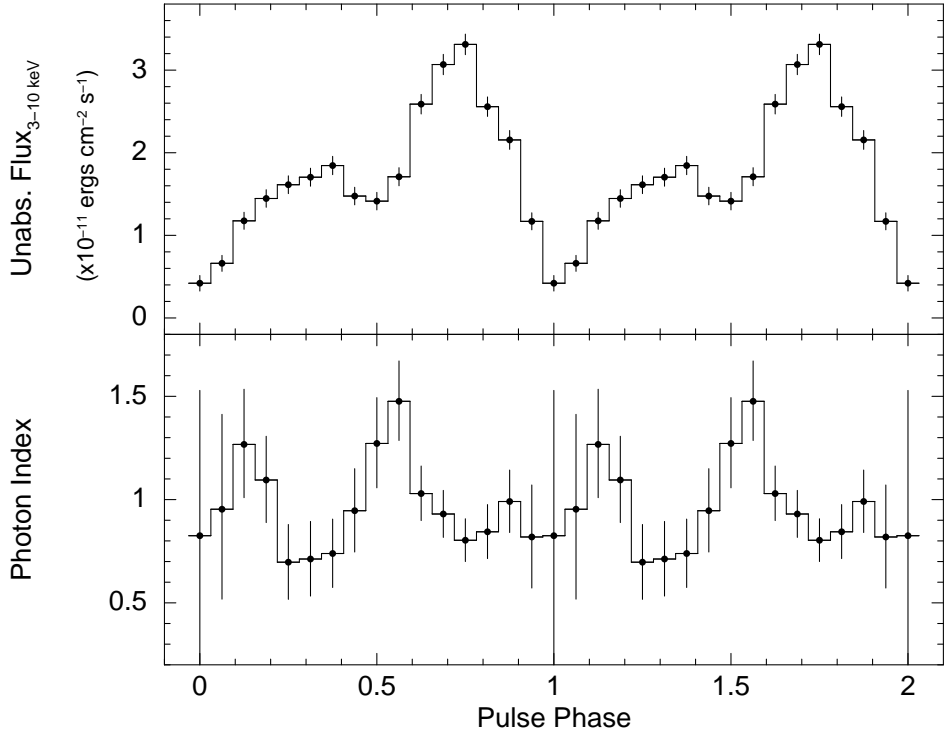}} 
   \caption{Pulse phase resolved spectral results deduced from 3--16.5 keV \emph{RXTE}$-$PCA spectra of 16 phase bins.
    3--10 keV unabsorbed flux values are plotted in the top panel. The photon index variation is plotted on the bottom panel (discussed in text). 
    All uncertainties are calculated at the 90 per cent 
   confidence level. For clarity, the data points are repeated for a cycle.}
   \label{phasespe}
 \end{figure}

In order to investigate the variation of spectral parameters with pulse phase, we analyze \emph{RXTE}$-$PCA observations 
between MJD 54935--54950 using the tool \verb"FASEBIN". Timing and orbital parameters for SWIFT J0513.4--6547 are appended 
into the timing files (i.e. \verb"psrtime.dat" and \verb"psrbin.dat") and 16 phase-binned spectra are constructed for the 
analysis. The exposure time of each bin turns out to be $\sim$1.1 ks. 3--16.5 keV spectra are modeled with the high energy 
cut-off model as described in Section 4.2.

In Fig. \ref{phasespe}, 3--10 keV unabsorbed flux and photon index values are plotted over pulse phase. The photon index 
exceeds its average value 0.93 in some certain phases. The steepest power law with index 1.48 is acquired in-between the 
double peaks of the pulse, whereas the peaks have lower indexes such as 0.7 and 0.8. The spectral parameters of lowest count 
rate bins (i.e. phase 0.0 and 0.0625) are not constrained well, an average index is found with a large uncertainty for these 
phases. 

\section{Discussion and Conclusion}

The X-ray luminosity of 2009 outburst was initially at its peak value of $L_{x} \sim 1.3 \times  10^{38} $erg s$^{-1}$ (Coe et al. 2015). It was found to decrease below $\sim 10^{37} $erg s$^{-1}$ towards the end of the outburst. The source disappeared for $\sim 2000$ days and then reappeared in 2014 with $L_{x} \sim 10^{36} $erg s$^{-1}$.

Our frequency measurement from 2009 observations which correspond to the final stage of the outburst (see Table \ref{timing}) complies with the general spin up trend of the initial stage of the outburst presented by Coe at al. (2015). Moreover, it is important to note that the arrival times bolster the Coe et al. (2015) assumption that the orbital and optical periods are the same.

The source is found to be spinning up with a rate of $1.74\times 10^{-11}$ Hz s$^{-1}$ during the final stage of the outburst of the 2009 outburst. Comparison with the values found for the initial stages of the 2009 outburst  ($\sim 3.75 \times 10^{-11} $Hz s$^{-1}$, Coe et al. 2015) suggests that the spin rate of the source is correlated with the X-ray luminosity.

On the other hand, from Table \ref{timing}, it is also found that the long-term ($\sim 1966$ days long) average spin-down rate of the source during the quiescence between 2009 and 2014 is  $\sim 1.52 \times 10^{-12}$ Hz s$^{-1}$ which is about two order higher than the rate measured during the re-brightening on 2014. This shows that, when the X-ray luminosity of the source increased in 2014, its spin-down rate is found to be smaller (i.e. spin rate value becomes greater).

Thus, whether the source spins up or spins down, its spin rate increases with the X-ray flux. Qualitatively, this trend suggests in general that the source spin rate is correlated with the X-ray luminosity. Observations from future outbursts of the source will be useful to better understand the spin rate $-$ X-ray luminosity relation of the source.

It is important to note that the long-term spin-down rate of $\sim 1.52 \times 10^{-12}$ Hz s$^{-1}$ corresponds to a state where the source is quiescent. Although this spin-down rate is comparable to the long-term spin-down rates of persistent accretion powered X-ray pulsars such as GX 1+4 (Gonzales-Galan et al. 2012), it is not typical within the quiescent states of the transient sources. Although some transient sources are either found to be spinning up or spinning down in their quiescent states in between outbursts (e.g. GS 0834-430, A 0535+262, see Bildsten et al. 1997), only a few of them such as SXP 7.78 (Klus et al. 2014) exhibits long-term ($\sim 1000$ days long) quiescent states for which the spin-down rates have the same order of magnitude. 

If we assume that the neutron star is in propeller state throughout the whole quiescent time interval, magnetic dipole ($\mu \approx BR^3$ where B is the surface dipole magnetic field and R is the radius of the neutron star) can be estimated (Alpar 2001) as

\begin{equation}
\label{eqn1}
\mu = {{(I|\dot{\Omega}|GM)^{1/2}} \over {\Omega}},
\end{equation}  

\noindent{where $I$ is the moment of inertia, $\dot{\Omega}$ is the spin rate, $M$ is the mass and $\Omega$ is the frequency of the neutron star. Using Equation \ref{eqn1}, for a neutron star of mass $1.4 M_{\odot}$, moment of inertia of $10^{45}$ g cm$^2$ and a radius of $10^6$ cm, surface dipole magnetic field is estimated to be $\sim 1.5 \times 10^{13}$ Gauss which is in good agreement with the magnetic field estimation by Coe et al. (2015) during the accretion phase using spin-up rate $-$ luminosity relation.}

We confirm that SWIFT J0513.4--6547 has a circular orbit by refining the upper limit for the orbital eccentricity as 0.028 in $2 \sigma$ confidence level. Although many of the Be/X-ray pulsar binary systems (especially the ones that have transient nature) have eccentric orbits, some of the transient Be/X-ray pulsar systems such as KS 1947+300 (Negueruela et al. 2003), 4U 1901+03 (Liu et al. 2003) , 2S 1553-542 (Kelley et al. 1983) and RX J0520.5-6932 (Kuehnel et al. 2014) have orbits with small eccentricities.

As seen in  Fig. \ref{enres}, it is evident that the pulse profile of SWIFT J0513.4--6547 is double-peaked. The count rate of one peak significantly exceeds the other, which we might call the primary peak. Regardless of the energy band, the peaks have similar phase coverage, where the mid-point between primary and secondary peaks is at $\sim0.5$ phase.

As seen in Fig. \ref{hard}, hardness ratios of the source correlates with the X-ray luminosity when 3-10 keV unabsorbed flux values are less than $2.8\times 10^{-11}$ ergs cm$^{-2}$ s$^{-1}$, while for higher flux values hardness ratios remain constant. With an assumed source distance of 50.6 kpc, the flux value separating the correlating hardness values and constant hardness values corresponds to a luminosity of $8.4\times 10^{36}$ erg s$^{-1}$. Postnov et. al. (2015) found similar hardness ratio dependence on X-ray luminosity in six transient X-ray pulsars (EXO 2030+375,GX 304-1, 4U 0115+63, V 0332+63, A 0535+26 and MXB 0656-072). Postnov et al. (2015) explains this "saturation" behaviour of the hardness ratio as a result of $\dot{M}$ surpassing a critical value  that leads to an increase in the height of the accretion so that the contribution of the reflected component to the total emission starts decreasing. But their corresponding critical luminosity is nearly one order higher than our value, being $\sim (3-7)\times 10^{37}$ erg s$^{-1}$.    

Maximum value of the pulse fraction obtained from 3-16.5 keV \emph{RXTE}$-$PCA observations (see Fig. \ref{pf}) is $\sim 0.8$ which is consistent with the previously reported value  (Krimm et al. 2009). In Fig. \ref{pf}, the pulsed fraction shows a clear correlation with the source flux, given that higher fractions are observed for higher flux. 
For higher accretion rates, the material pressure of the accreting mass allows the plasma to move in and induces a closer 
interaction between the plasma and the magnetosphere. Consequently, increased efficiency of accretion onto the magnetic 
poles results in a higher pulsed fraction.

We found that overall \emph{Swift}$-$XRT and \emph{RXTE}$-$PCA spectrum of SWIFT J0513.4--6547 is well described with either a power law model with a high energy cut-off at $\sim 5$ keV or a model consisting of a black body peaking at $\sim 1.9$ keV and a power law. For the overall spectrum, both of these models fit equally well. 

If the model consisting of a blackbody component is considered; not only a quite high value of blackbody temperature ($\sim 1.9$ keV), but also the unusually high value of blackbody emitting radius ($\sim 1.7$ km) makes the model not favorable. Usually such high kT's are observed in the spectra of low luminosity ($\sim 10^{34}$ erg s$^{-1}$) accreting X-ray pulsars, although there might be exceptions such as SAX J 2103.5+4545 and SWIFT J045106.8-694803 with a source luminosity and black body kT of the order of those of SWIFT J0513.4--6547 (Bartlett et al. 2013; \.{I}nam et al. 2004a). Moreover the fraction of blackbody flux 
contributing to the 0.3-10 keV flux is rather low as about 30 per cent of the unabsorbed flux. On the other hand, if the model including a power law with high energy cut-off is considered,
high energy cut-off is a usual spectral feature that is observed in many of the HMXBs. Even if the cut-off energy 
$E_{\mathrm{cut}}=5.04^{+0.77}_{-0.67}$ keV measured for SWIFT J0513.4--6547 is at the lowest end of the range of energies measured 
for other sources (e.g. for 4U 1626--67 $E_{\mathrm{cut}}=6.8$ keV, see Coburn et al. 2002), we still prefer to represent the individual spectra with this model.

From Fig. \ref{spe2009}, there is a marginal evidence that the decrease in X-ray flux leads to a softening of the X-ray spectrum. Similar spectral dependence which might be an indication of accretion geometry changes due to mass accretion rate variations was also found for the X-ray pulsars SAX J2103.5+4545 (Baykal et al. 2002), 2S 1417-62 (\.{I}nam et al. 2004b), XMMU J054134.7-682550 (\.{I}nam et al. 2009), Her X-1 (Klochkov et al. 2011) and SWIFT J1626.6-5156 (\.{I}cdem, \.{I}nam and Baykal 2011).  

We also studied pulse phase resolved spectra of the source using \emph{RXTE}$-$PCA observations of 2009. As seen from Fig. \ref{phasespe}, the peaks have similar photon index values of $\sim0.7-0.8$. However, the midpoint between primary and secondary peaks at $\sim0.5$ phase has a steeper (softer) spectrum with a power law index of 1.48. The softening for this phase is also evident in HR1 hardness ratio shown in Fig. \ref{hrp}. The hardness ratio involving higher energy bands (i.e. HR2) reveals that the spectrum of the primary peak is softer than that of the secondary peak.   This may be due to the fact that the higher energy bands have less count rates, so even modest variations in these energy bands do not affect the power law indices significantly.

\section*{Acknowledgment}

We acknowledge support from T\"{U}B\.{I}TAK, the Scientific and Technological Research Council of Turkey through the 
research projects TBAG 109T748 and MFAG 114F345.

\bsp

\label{lastpage}

\end{document}